\documentstyle[preprint,pra,aps]{revtex}
\begin{document}
\draft
\title{ Dirac-Fock calculation for H, H$_2^+$ and H$_2$ in a strong \\
magnetic field by the Hermitian basis of B-splines}
\author{G.B.Deineka}
\address{Department of Physical Chemistry, Fiber and Integrated Optics.\\
St.Petersburg State Institute of Fine Mechanics and Optics\\
(Technical University),14 Sablinskaya, 197101 St.Petersburg, Russia}
\date{\today}
\maketitle

\begin{abstract}
A two-dimensional, fully numerical approach to the solution of
four-component Dirac-Fock equation using the moderately long
Hermitian basis of B-splines is applied to H, H$_2^+$ and H$_2$ in a
strong magnetic field.
The geometric parameters, including
different behavior of wave-functions relativistic components are analyzed.
The accuracy of the solutions as a function of the basis lenght is
estimated. The relativistic corrections are calculated by transformation
of the  matrix equations  to the equations  for large relativistic
components.
Application of the finite-element method to solution of the
Dirac-Fock equation without supplementary assumption about exchange in
case of the H$_2$ excited states is discussed.

The maximum localization
of the basis functions provides applicability of the quadrature formulae
for five-dimensional two-electron integral calculations within
reasonable period.
\end{abstract}


\narrowtext

\section{ Introduction}
\label{sec:1}

The problem of the hydrogen atom and hydrogen molecular ion in the strong
magnetic field has wide application in the investigations of "white
dwarfs" spectral properties in astrophysics.  Many numerical and
analytical calculations have been reported in  literature
\cite{C1,C2,C3,C4,C5,C6,C5.1,C7}.
The moderately low temperature (5000K - 25000K) of
the "white dwarfs" \cite{C8} make reasonable the consideration not only
of the
one-electron system (H, H$_2^+$), but also of the hydrogen molecule
\cite{C5.1,C5.2,C5.3}.
Most
of the previous works treat the effect of strong ($\approx10^9$ G) magnetic
fields on the electron structure in the framework of non-relativistic
theory, but in  case of the  magnetic field magnitude is comparable with
that of the light velocity (~137 in a.u.), the relativistic  approach is
then more consistent.
First and foremost because the relativistic one describes the features
of the particle with spin in electromagnetic field more correctly.
In present paper we neglect  the proton motion in magnetic field.
Movement of this type may have an expressed effect on energetic structure
in case of superstrong($ > \approx10^{12}$ G) magnetic field ( it
was estimated i.c. in \cite{C4.1} ). We aware of the fact  that the
relativistic correction  in our case may be comparable to effect
of the nuclear motion and this problem calls for further investigation.

The one-electron system of H was studied in  framework
of relativistic theory in  detail  with high accuracy
(see Ref. \cite{C4} ).
The molecular systems H$_2^+$ and H$_2$ in the
strong magnetic field  were studied, as a rule, as a non-relativistic
problem (see recent papers ref. \cite{C5} for H$_2^+$ and
\cite{C5.2,C5.3,C5.4} for H$_2$ ).  In the field-free relativistic
case the maximum accuracy (better then $10^{-10}$ au) was obtained
with use of  the finite
element method \cite{C11,C15}. The authors  used the Lagrangian-type
finite elements defined as two-dimensional fifth(and more)-order polynomials
within triangular regions \cite{C10}. But for the molecular system
with the exchange the cited authors would have to use
supplementary Dirac-Fock-Slater approximation \cite{C11}.
From our viewpoint the reason is that the Lagrangian-type finite elements
are not enough localized for  calculations of electron interaction matrix
elements in a reasonable time (e.g. the fifth-order polynomial occupies
21 nodal points).
As it was shown in \cite{C5.4} the state  $^3\sum_u$ with parallel
spin becomes the ground state of the hydrogen molecule in magnetic
field of strengths $B/B_0  > 0.2$($B_0\,=\,2.35\times10^9 G$).
I.e. the exchange interaction must be taken into account.

The main difficulty in relativistic calculations
consists in a generalized  matrix problem enlarged by four times.
Furthermore the energy matrix with an exchange loses the band structure
and the time of calculation mainly depends on the exchange matrix
construction.
For the solution of exchange problem in a reasonable time we have used the
B-splines basis set with  minimum overlap.
Taking into  account the high accuracy of up-to-date relativistic
solutions for H in the magnetic field  and  for H$_2^+$  in free-field
case, we can estimate the capabilities of the finite function basis
with maximum localization. The results of this work have shown that
even with the moderate length it is possible to use this basis for
the successful calculations of the energy relativistic corrections
and geometric parameters of H, H$_2^+$ and H$_2$ in  strong
magnetic field.
Following Z.Chen and S.P.Goldman \cite{C4} we have noted that
variation method in relativistic case "may have difficulties of
variational collapse, spurious roots, continuum dissolution in case
of system with more then one electron" (in the  one-electron system
the wave functions with  poor quality may have the  lower-lying
state energy, than that from the exact solution of the problem).
To avoid spurious roots in case of H$_2$ we used the direct
QL-algorithm for finding all of the generalized eigenvalue problem
eigenvectors \cite{C14}.
Since we investigated the different behavior of one-electron wave-function
relativistic components in the magnetic field, the calculation
of H$_2$ was performed only in the framework  of the one-determinant
approximation.

By these means in
this work we realize a two-dimensional, fully numerical approach to the
solution of the four-component Dirac-Fock equation using the Hermitian-type
basis of B-splines \cite{C9}. We use the Hermitian
basis set with  maximum localization of each finite element (i.e.
the overlap exists only between the nearest neighbors).
In addition, the
Hermitian basis sets provide the fast expansion and re-expansion of
wave function only with  use of  function values and their
derivatives in every nodal point\cite{C12}. These features of the basis set
allow to calculate the Dirac-Fock equation without  supplementary
assumption of  exchange \cite{C9.1} in the case of the H$_2$
states in the magnetic field.

In this paper we have compared the energy levels of the hydrogen atom and
the hydrogen molecular ion with the corresponding values in
relativistic and non-relativistic calculations, have demonstrated
the difference in
behavior between the large and small relativistic components under
action of  magnetic field and have estimated the magnitude of  magnetic
field for transitions in the hydrogen molecule in the framework of
one-determinant approach.

\section{Dirac-Fock finite-element calculation}
\label{sec:2}

The finite-element method (FEM) is a well established
technique for solution of several quantum mechanical
problems. Recently, it has been demonstrated \cite{C10,C11}  that
this method can also be used for solving the
many-electron Dirac problem.  The finite basis functions
presented in these publications have the following
features. First, each finite function (FF) occupies more
space than a region between the main function nodal point and
the nearest nodal points. Second, the increasing of
accuracy is provided, as a rule, by increasing the nodal
points number for every finite function. The insufficient
localization of FF impedes the calculation of two-electron
matrix elements, therefore usually  \cite{C10,C11} a Slater's local
approximation  for exchange potential in diatomic molecules
calculations  is used.  In this work we present  a FF basis
set  with  maximum localization for the two-center
many-electron Dirac problem. This set we have constructed
from Hermitian basis of B-splines (HBS) at every nodal
point.  The HBS analytical  properties  and their
application  to the solution of the one-center Fock-Dirac
equations we have discussed in the above Ref.\cite{C9}. The relativistic
Hamiltonian operator for a  diatomic molecule
electron subsystem is defined as

\begin{equation}
H\,=\,
\sum_{j}\,\Bigl[\,c\,{\bbox \alpha}_j\,({\bf p}_j\,-\,
i\,{{\bf A}\over{c}})\,
+\,c^2\beta_j\,-\,{{Z_1}\over{r_{1j}}}\,-\,{{Z_2}\over{r_{2j}}}\Bigr]\,+\,
\sum_{i,j}\,{1\over{r_{ij}}},
\end{equation}
where  ${\bbox \alpha}$ and $\beta$ are usual Dirac matrices,
Z$_1$ and Z$_2$ are the
nuclear charges, r$_1$ and r$_2$ are the distances to the nuclei,
${\bf A}\,=\,(-By,Bx,0)$ is the vector-potential of the  electromagnetic
field, $B$  is the magnetic field magnitude in  z-direction.
Because of  axial symmetry of diatomic molecules, we
describe the four-component spinor of a single electron
wave function $\Phi$ in elliptic  coordinates  ($\rho$,$\tau$,$\varphi$):

\begin{eqnarray}
\rho\,=\,{{r_1+r_2}\over{D}},\nonumber\\
\tau\,=\,{{r_1-r_2}\over{D}},\nonumber
\end{eqnarray}
where $D$ - is the distance between  nuclei.
After  separation  of the angular coordinate $\varphi$  and
modification of ($\rho$,$\tau$) coordinates by the introduction of
the parameters for  the coordinates "stretching"  ($d_{\rho}$ and
$d_{\tau}$)

\begin{eqnarray}
s(\rho)\,=\, {{\rho^2}\over{\rho\,-\,(1-d_{\rho})}},\nonumber\\
t(\tau)\,=\,{{\tau}\over{(1+d_{\tau})^2\,-\,\tau^2}}
\label{st}
\end{eqnarray}
we obtain for  single electron  wave function within
z-projection of total momentum $\mu\,$ the following  expression:

\begin{eqnarray}
\Phi\,=\,
 \left(
\begin{array}{c}
\phi_0(s,t)\, e^{i(\mu-1/2)\varphi}\\
\phi_1(s,t)\, e^{i(\mu+1/2)\varphi}\\
\phi_2(s,t)\, e^{i(\mu-1/2)\varphi}\\
\phi_3(s,t)\, e^{i(\mu+1/2)\varphi}\\
\end{array}
\right).
\label{Fi}
\end{eqnarray}

The relativistic components $\phi_k$ are expanded in terms of the FF:

\begin{equation}
\phi_k(s,t)\,=\,
[(\rho^2-1)\,(1-\tau^2)]^{(\mu-1/2+\lambda)/2}\,
 \sum_{i_s,i_t,h_s,h_t}\,
C(i_s,i_t,h_s,h_t,k)\,
 F(i_s,i_t,h_s,h_t;s,t),\, \\
\label{ä=c*f}
\end{equation}
where $(i_s=1..N_s, i_t=1..N_t)$ are the nodal indices for $(s, t)$
coordinates, $(h_s, h_t)$ are the Hermit component indices and
$\lambda$=(0,1,0,1) for $k$=(0,1,2,3).
In our work the two-dimensional FF is constructed as a
product of the one-dimensional elements from the HBS $f(i,h;x)$:

\begin{equation}
F(i_s,i_t,h_s,h_t;s,t)\,=\,f(i_s,h_s;s)\,f(i_t,h_t;t)\,\\
\label{F=f*f}
\end{equation}

The analytical properties of two first Hermit components
of the B-splines $f(i,h;x)$  are the following:

\begin{eqnarray}
f(i,0;x)\,=\,
\left\{
\begin{array}{c}
0,\,\bigl\vert\,a\,\bigr\vert\,\ge\,1\,\\
(1-a)^2\,(1+2a),\,a\,\ge\,0,\\
(1+a)^2\,(1-2a),\,a\,<\,0,\\
\end{array}
\right.
\nonumber\end{eqnarray}
\begin{eqnarray}
f(i,1;x)\,=\,
 \left\{
\begin{array}{c}
0,\,\bigl\vert\,a\,\bigr\vert\,\ge\,1\,\\
(1-a)^2\,a,\,a\,\ge\,0,\\
(1+a)^2\,a,\,a\,<\,0,\\
\end{array}
\right.
\label{B-s}
\end{eqnarray}
where $a\,=\,(x-i\,d)\,/\,d$, $\,d\,$ is  the distances between the
nearest nodal points.

Therefore, each FF
occupies the region between the main function nodal point
$(i_s,i_t)$ and the nearest nodal points $(i_s\pm\,1,i_t\pm\,1)$ only.
We shall call this region the FF local space.  We have calculated
the $H$ matrix
elements in the crossing of the FF local
spaces numerically, using the seven-point  quadrature
formula (these formulae are exact for the polynomials
with the degree 6 and under). We have used the equi-distant net
with 24 points of integration between the nodal points. The
expansion (\ref{ä=c*f}) leads to a $(16\,N_s\,N_t)\times(16\,N_s\,N_t)$
symmetric  eigenvalue self-consistent-field equation for
the one-electron vector $C$
\begin{equation}
(H_m\,+\, J\,-\, K)\, C\,=\,E\,S\,C\,,
\label{hc=esc}
\end{equation}
where $H_m,\, J,\,\, K\,$ and $\,S\,$ are the one-electron, Colomb's,
exchange and overlap matrix operator, respectively.
The number $m\,=\,\mu\,-\,1/2\,$ defines the symmetry properties of the
wave function $\Phi$. The choice of the two-dimension FF as a
product of one-dimensional ones allows  to reduce most
of one-electron matrix elements to the one-dimension
integrals. In case of the two-electron elements we used the
interpolation properties of the Hermitian-type
$FF$ \cite{C12}

\begin{equation}
Z(x,y)\,\approx\,\sum_{i_x,i_y,h_x,h_y}\, C(i_x,i_y,h_x,h_y)\,
f(i_x,h_x;x)\,f(i_y,h_y;y),
\label{Z}
\end{equation}
with  the nodal values
\begin{equation}
C(i_x,i_y,h_x,h_y)\,=\,\bigl({\partial \over \partial{x}}\bigr)^{h_x}\,
\bigl({\partial \over \partial{y}}\bigr)^{h_y}\,
Z(x,y)\bigr\vert_{i_x,i_y},
\label{C}
\end{equation}
i.e. by the approximation of an arbitrary differentiable
function the nodal values C are defined by the function
and  derivative values in the nodal points.

Let us
consider, for example, the exchange matrix element for
the single-determinant case (in Matsuoka's \cite{C13} notations):
\begin{equation}
K_{\alpha,\alpha1}^{k,k1}\,=\,<\,\alpha\,\vert\,<\,
\phi_{k1}\vert{\,g(s,t,s_1,t_1)\,}\vert\,\alpha1\,>\,\vert\,
\phi_{k}\,>\, =\, < \,\alpha\,\vert\, G_{\alpha1,k1}(s,t)\,\vert\,
\phi_{k}\,>\,,
\nonumber
\end{equation}

where  $\alpha\,=\,(i_s,i_t,h_s,h_t,k)$  is the generalized FF index.
Function $\,G_{\alpha1,k1}(s,t)\,$
is the result of the integration over a relatively
small-sized region of the FF $\,F(\alpha1;s,t)\,$ non-zero values.
Moreover; in view of Eq.(\ref{Z}) and (\ref{C}), the function
$\,G_{\alpha1,k1}(s,t)\,$ and
its derivation can be calculated for the nodal points
$(i_s, i_t)$ only.
In the end of this section we present  our results for H$_2^+$ ground
state in the field-free case (third-order piecewise polynomials (\ref{B-s}))
in comparison with relativistic calculation on the base of fifth-order
polynomials (Table \ref{tableA}).
 The "exact" values in Table \ref{tableA} for the relativistic orbital
energies
are obtained in the calculation of L.Yang et al. \cite{C15}  by adding the
relativistic energy correction  to the nonrelativistic energy
of Madsen and Peek \cite{C16}.
In Table \ref{tableB} are presented the values of the ground state energy
for $H$ and $H_2$ with the same parameters of FF (exept internuclear distance)
as for $H_2^+$  in Table \ref{tableA}. The energy values for $H$ are
compared with analytical Dirac equation solution  for Coulomb field.
The calculation accuracy  of $H$ and $H_2^+$ are closely  allied, the
accuracy for $H_2$ is one order of magnitude less then for one-electron
case.
The relativistic corrections are defined by  nonrelativistic limit
of (\ref{hc=esc}) (a matrix equation  for large relativistic components).
The accuracy of relativistic corrections for  $H$,  $H_2^+$ and  $H_2$
in the field-free case  are in good agreement  with the results of
special-purpose relativistic FF calculations.
The values of the  relativistic corrections is calculated
as a difference ( result of  subsraction) between relativistic value
and value in non-relativistic limit (for matrix one-electron Fock-Dirac
equation (\ref{hc=esc})) in the same finite functions basis.
The aproach of this kind allows to compensate (in great part)
in the relativistic corrections   uniform errors,
associated with the basis lenght, and to obtain the
reasonable  values with using the moderately long
B-splines basis.   This correction (as it is
shown in Tables \ref{tableA},\ref{tableB} for field-free case and
\ref{table1}  for H in magnetic field) good fit to the results,
obtained  with the energy high-precision calculation.

\section{Results}
\label{sec:3}

In the one-electron case, the equation (\ref{hc=esc}) was solved for
the ground states of the hydrogen atom and hydrogen
molecular ion. The nuclei were placed into the coordinate
system focal points.  In the coordinates (\ref{st}),
the $N_s\times N_t$ equi-distant
nodal net was built for the finite functions $F\,$ from the expansion
(\ref{ä=c*f}). The parameters $d\rho\,$ and $d\tau\,$ were chosen with
use of the
optimization for field-free case.  The $13\times9\,$ for $H$ and
$10\times11\,$ for $H_2^+$ nodal net with two
Hermitian components of B-splines (\ref{B-s}) and four relativistic
components of wave functions leads correspondingly to the
$1872\times1872\,$  and  $1760\times1760\,$ generalized matrix problem and to the
468 and 440 basis function per one relativistic component
(See $N_b$ in Table \ref{tableA} and Table \ref{tableB} for
the accuracy estimation).
This problem was solved for all of the
eigenvalues by the method from Wilkinson and Reinsch handbook
\cite{C14}. It takes about 150 min for $H$ or, if we take into account
the system symmetry,  45 min for $H_2^+$ of CPU time on a P-II-233 IBM
compatible computer. The equilibrium internuclear distance
was calculated with use of the quadratic approximation or with
use of relativistic virial theorem for diatomic molecule in magnetic field
(the specific cases of relativistic virial therem for atoms was considered
in \cite{C4,C17}).
Both of these methods lead to the near results. The results
are listed in Tables \ref{table1} and \ref{table2}.
For comparison with other
results, we have used as  the convention the parameter
$B/B_0$, with $B_0\,=\,2.35\times10^9 G$ (137 a.u.).

Our results for $H$ presented in Table \ref{table1} are
compared with  detailed data from the non-relativistic
B-splines calculations made by Jinhua Xi et al.\cite{C7} and with
relativistic finite basis set calculations by
Z.Chen and S.P.Goldman \cite{C4}. For the
field values ($B/B_0$ from 0 to 10) Jinhua Xi et al. have used
$35\times16\,=\,560\,$ nodal net, that is by 1.2
times more than for the separated relativistic component in our
calculations (468).
Although in our calculations the accuracy of energy is less then in
\cite{C4}, the magnitudes of relative relativistic corrections
are in good agreement  with the results of Z.Chen and S.P.Goldman.
Table \ref{table1} shows the magnetic field compression
effect on the  relativistic wave function too. The  average
radius in x-y-direction $R_{xy}$ (i.e. average radius in cylindrical
coordinat system) decreases monotonical both for the
total density and for the small component (k=2,3 in (\ref{Fi}) ).
But for the small component, one decreases to a greater extent
for $B/B_0\,$ values from 0 to 20.0.
The average length in z-direction $R_z$ (root-mean-square deviation of
z-projection vector from nucleus) for the total density decreases
to a greater degree then the small component.

In the Table \ref{table2}  our results for the ground state
of H$_2^+$ are presented. The  symmetrical   state     $1(-1/2_g)^1$
in Herzberg's notation for  double group symmetry (See \cite{C11,C13.1})
is considered.

In case of antiparallel orientation of the field and  z-projection of total
moment (positive $B/B_0$)
the relativistic correction, transversal size of the ion  electronic
density and size of small component decrease under effect of the magnetic
field in the same manner as for $H$.

The reverse statement is true for parallel orientation (negative $B/B_0$).
The relativistic correction quadratically increases by increasing of
magnetic field intensity. The similar behaviour can be recognized for
$H$ in Table \ref{table1} (marked with [$^c$]).
The transversal size $R_{xy}$ of small component increases by changing
$B/B_0$ from 0 to about -0.2, restores its original dimension
at $B/B_0\approx-1.0$ and decreases at $B/B_0 < -1.0$.

In the Table \ref{table3} we present the "unrestricted" Dirac-Fock
( different orbitals for different z-projection of total momentum)
 results for the singlet
state of H$_2$ with the relativistic configuration $1(1/2_g)^2$.
The state energy has the local minimum as the
function of the internuclear distance for all of the considered field
magnitudes.  The equilibrium internuclear distance  was calculated
by the quadratic approximation in the vicinity of the minimum.
The total longitudinal and transversal sizes of the hydrogen molecule
decreases, as well as for the hydrogen ion, but the field dependence
of the small components is different for electrons with $\mu\,=\,-1/2$ and
$\mu\,=\,+1/2\,$ ($\mu\,$ is z-projection of total one-electron momentum).
Average  radius of the small component in x-y-direction for the
electron with $\mu\,=\,-1/2\,$ decreases under the field compression effect
monotonously. By contrast, for the electron with $\mu\,=\,+1/2\,$ one
increases up to the field magnitude $B/B_0\,=\,0.4\,$ and remains greater
than the radius in the field-free case up to $B/B_0\,=\,1.0\,$. The
same behavior of the small component is to be observed for the
$1(-1/2_g)$ and $1(+1/2_g)$  configurations of H$_2^+$
($\pm B/B_0$ in Table \ref{table2}).

The relativistic correction of total energy increases quadratically in
the same manner as for $H_2^+$ in the case of parallel orientation.

Comparison between total energies for H$_2$ and  the sum of two energies
for hydrogen atoms (see Table \ref{table1}) shows that at $B/B_0\,>\,0.2\,$ the state
with two insulated atoms has the total energy less than that of the molecule.
For better visualization of  dissociation channel  of H$_2\,$ into the
single atoms, we have considered the energies of the nearest
configuration ($1(-1/2_g)1(-1/2_u)\,\,$)
with no change of the equilibrium
internuclear distance.  The energy of this configuration lie in the
field-free case well above the ground state. With increasing of the
field magnitude, this energy decreases and become equal to the
$1(1/2)^2$ configuration
energy in the vicinities $B/B_0\,\approx\,0.35 ... 0.4\,$.
It is nessesary to stress that we have used the one-determinant
approximation.
In the non-relativistic configuration-interaction calculation
by T.Detmer at al. \cite{C5.4} the $^1\sum_g$ and $^3\sum_u$ states cross
over at $B/B_0 \approx > 0.2$.
When taken  into account Detmer's total energy of the $^1\sum_g$ state at
$B/B_0 = 0.2$ (-1.158766a.u.)  and total energy  of two unsulated $H$
atoms by Jinhua Xi et al.\cite{C7} (-1.18076313 a.u.) the cross point
is found to be at $B/B_0 < 0.2$.

The energy
of the $1(-1/2_g)1(-1/2_u)\,$ state is free of minimum \cite{C5.4}
as being an internuclear distance function.
And so, in the case of
transition into $1(-1/2_g)1(-1/2_u)\,$ configuration, the hydrogen
molecule dissociates into  single atoms.

As it is shown in Table \ref{table3.1} the  z-direction compression
coefficient for $H$ has given better fit to the relative reduction of
internuclear distance  in  $H_2^+$ and $H_2$ than x-y compression one.

\section*{Acknowledgements}

This work has been implemented under financial support of
the Russian Foundation
of Fundamental Investigations (Grant No.  96-03-33903a). The author would
like to thank Professor I.V.Abarenkov and Dr. I.I.Tupitsin for their helpful
comments and encouragement. The author is grateful to Professor Elizabeth
Davis for reading the manuscript.

\begin{table}
\caption{ Hydrogen ion ground state energies (Es), accuracy of Es,
relativistic energy corrections(Rc) with  increasing number of basic
function per one relativistic component (Nb in our calculation )
or number of grid points (Np by L.Yang et al.). The values are given in au.
Internuclear distance for $H_2^+$ is 2.0 a.u.}
\label{tableA}
\begin{tabular}{dcccdccc}
Nb&Es&Accur.Es&Rc&Np&Es&Accur.Es&Rc \\
& & &$\times10^{-6}$& & & &$\times10^{-6}$ \\
&\multicolumn{4}{c}{Present work         }&{L.Yang et al. \cite{C15} }\\
\tableline
120&-1.1027204866  &7.9$\times10^{-5}$&-7.3689  & 121 &-1.10408781      &1.3$\times10^{-3}$&-7.22      \\
168&-1.1026813906  &4.0$\times10^{-5}$&-7.3667  &     &                 &       &          \\
224&-1.1026596742  &1.8$\times10^{-5}$&-7.3667  &     &                 &       &          \\
288&-1.1026512414  &9.6$\times10^{-6}$&-7.3666  & 256 &-1.102694840     &5.3$\times10^{-5}$&-7.358     \\
360&-1.1026473438  &5.7$\times10^{-6}$&-7.3666  &     &                 &       &          \\
440&-1.1026452809  &3.7$\times10^{-6}$&-7.3666  & 441 &-1.1026462146    &4.6$\times10^{-6}$&-7.3660    \\
528&-1.1026440929  &2.5$\times10^{-6}$&-7.3665  &     &                 &       &          \\
624&-1.1026433680  &1.8$\times10^{-6}$&-7.3665  & 676 &-1.10264199741   &4.0$\times10^{-7}$&-7.36651   \\
   &               &      &        & 961 &-1.102641606652  &2.6$\times10^{-8}$&-7.366544  \\
   &               &      &        &2601 &-1.10264158113086&1.0$\times10^{-10}$&-7.36653872\\
   &               &      &        &exact
\tablenote{ See text }
&-1.1026415810336 &       &-7.3665387 \\

\end{tabular}
\end{table}

\begin{table}
\caption{$H$ and $H_2$ ground state energies(Es), accuracy of Es,
relativistic  corrections(Rc) with increasing number of basic
function per one relativistic component(Nb). Internuclear distance
for $H_2$ is 1.4 a.u.
The values are given in a.u.}
\label{tableB}
\begin{tabular}{ccccccd}
Nb&Es&Accur.Es&Rc&Es&Accur.Es&Rc \\
& & &$\times10^{-6}$& & &$\times10^{-6}$  \\
\tableline
&$H$& & &$H_2$ & &  \\
\tableline
120&  -0.500098633 &9.2$\times10^{-5}$&-6.664&
-1.13388849713&2.4$\times10^{-4}$&-14.401  \\
168&  -0.500079978 &7.3$\times10^{-5}$&-6.660&
-1.13387761541&2.4$\times10^{-4}$&-14.396  \\
224&  -0.500044313 &3.8$\times10^{-5}$&-6.658&
-1.13379246622&1.5$\times10^{-4}$  \\
288&  -0.5000247572&1.8$\times10^{-5}$&-6.6575&
-1.13372542209&8.1$\times10^{-5}$&-14.3975\\
360&  -0.5000167610&1.0$\times10^{-5}$&-6.6572&
  &  \\
440&  -0.5000128771&6.2$\times10^{-6}$&-6.6570&
-1.13367772292&3.4$\times10^{-5}$&-14.3984  \\
528&  -0.5000106448&3.3$\times10^{-6}$&6.6569 &
 -1.13366931610&2.5$\times10^{-5}$&-14.3986  \\
2116  \tablenotemark[1]
 & & &  &-1.133643970950&$<10^{-10}$&-14.399093\\
24897 \tablenotemark[2]
 & & &  &-1.133643970   &  &-14.397  \\
analyt.&-0.50000665659718&     &-6.656597 \\

\end{tabular}
\tablenotetext[1]{ See \cite{C11} }
\tablenotetext[2]{ See \cite{C11.1} }

\end{table}

\begin{table}
\caption{ Hydrogen atom ground state energies (Es),
relativistic correction (Rc), relative correction  ($R_c/E_{nrl}$),
compression
magnitude (K) for: - average
radius in x-y-direction for the total (Rxy) density and
small (Qxy) relativistic component;
-  dimension in z-direction for the total (Rz) density and
small (Qz) relativistic component.
In the calculations we have used 468  of basis function
per one relativistic component.
 The values are given in a.u.;
nr -  nonrelativistic calculations,
nrl - present work nonrelativistic limit.
}
\label{table1}
\begin{tabular}{cccccccc}
$B/B_0$&Es&$R_c/E_{nrl}$ & $R_c$ &K(Rxy)& K(Qxy)&K(Rz)&K(Qz)  \\
\tableline
 0.0&-0.5000066415  &13.31 &-6.657 &1.0(1.17777)&1.0(1.17811)&1.0(0.99959)& 1.0(1.81014)\\
 0.0&-0.4999999843  & nrl   &                               \\
 0.1&-0.5475340019  &10.82 &-5.928 &0.9914& 0.9484 & 0.9955& 0.9987\\
 0.1&-0.5475280736  & nrl   &                               \\
 0.1\tablenotemark[1]&-0.5475324083429& 10.8  &              \\
 0.2&-0.5903904145  &9.222 &-5.445 &0.9694& 0.8985 & 0.9839& 0.9958\\
 0.2&-0.5903849695  & nrl   &                               \\
 0.2\tablenotemark[2]&-0.5903815      & nr    &              \\
-1.0\tablenotemark[3]
&0.16878489816  &       &-40.112  &0.7552& 1.0057 & 0.8632& 0.8856\\
 1.0&-0.8311793578  &5.208 &-4.329 &0.7552& 0.6601 & 0.8632& 0.9587\\
 1.0&-0.8311750285  &nrl    &                               \\
 1.0\tablenotemark[2]&-0.831169       &nr     &              \\
 1.0\tablenotemark[1]&-0.831173226    & 5.21  &              \\
 2.0&-1.0222267871  &4.031 &-4.121 &0.6072& 0.5336 & 0.7716& 0.9114\\
 2.0&-1.0222226658  &nrl    &                               \\
 2.0\tablenotemark[2]&-1.022214       &nr     &              \\
 2.0\tablenotemark[1]&-0.022218029    & 4.03  &              \\
 3.0&-1.1645485075  &3.476 &-4.048 &0.5222& 0.4634 & 0.7150& 0.8738\\
 3.0&-1.1645444592  & nrl   &                               \\
 3.0\tablenotemark[2]&-1.164533       & nr    &              \\
 3.0\tablenotemark[1]&-1.164537038    & 3.48  &              \\
 4.0&-1.2808163924  &3.117 &-3.992 &0.4656& 0.4166 & 0.6752& 0.8440\\
 4.0&-1.2808124002  & nrl   &                               \\
 4.0\tablenotemark[2]&-1.280798       & nr    &              \\
 8.0&-1.6194194786  & 2.289&-3.707&0.3463& 0.3167 & 0.5845& 0.7680\\
 8.0&-1.6194157712  & nrl   &                               \\
 8.0\tablenotemark[2]&-1.619384       & nr    &              \\
10.0&-1.7478463122  & 2.015&-3.523&0.3134& 0.2886 & 0.5574& 0.7437\\
10.0&-1.7478427892  & nrl   &                               \\
10.0\tablenotemark[2]&-1.7477965      & nr    &              \\
20.0&-2.2157351467  & 1.077&-2.388 &0.2278& 0.2140 & 0.4798& 0.6714\\
20.0&-2.2157327590  & nrl   &                               \\
20.0\tablenotemark[1]&-2.215400913    & 1.09  &               \\
\end{tabular}
\tablenotetext[1]{ Z.Chen and S.P.Goldman \cite{C4} }
\tablenotetext[2]{ Jinhua Xi et al.\cite{C7} }
\tablenotetext[3]{ See comment to Table \ref{table2} }

\end{table}


\begin{table}
\caption{$H_2^+$ ground state electronic energies
(Es), relativistic correction (Rc), total energies(Et),
equilibrium internuclear distance(Re), compression magnitude (K) of
average radius in x-y-direction for the total (Rxy) density and small (Qxy)
relativistic component.
In the calculations we have used 440  of basis function
per one relativistic component.
All values are given in au;
nr -  non-relativistic calculations,
nrl - present work non-relativistic limit.}
\label{table2}
\begin{tabular}{ccccccc}
$B/B_0$\tablenotemark[1]
       & Es& Rc &Et&  Re& K(Rxy)& K(Qxy)\\
       &   &$\times10^{-6}$    &  &      &       &       \\
\tableline
 0.00\tablenotemark[2]
&           & nr        &-0.60263462&1.99719&       &       \\
0.00&-1.10334036& nrl   &-0.60263788&1.997194&(0.957659)
\tablenotemark[4]
&   \\
0.00&-1.10334774&-7.3772 &-0.60264525&1.997194&1.0(0.95764)
&1.0(1.08362)\\
0.02\tablenotemark[2]
&                &    nr&-0.61257052&1.99699&       &         \\
0.02&-1.113327412& nrl  &-0.61257377&1.99697&        &         \\
0.02&-1.113334597&-7.1850&-0.61258096&1.99697&0.99985 &0.99449  \\
0.10\tablenotemark[2]
&                & nr   &-0.65103820&1.99221&       &          \\
 0.10&-1.15299661& nrl  &-0.65104149&1.99221&0.99647 &         \\
 0.10&-1.15300313&-6.5229&-0.65104801&1.99221&0.99646 &0.97084 \\
-0.10&-1.05300513&-8.5243&-0.55105001&       &0.99647 &1.02410 \\

 0.20&-1.20187690&-5.8942&-0.69633157&1.97806&0.98651 &0.93907 \\
-0.20&-1.00188103&-10.027&-0.49633570&       &0.98652 &1.04133 \\

 1.00\tablenotemark[3]
&                &nr    &-0.97321   & 1.76    &       &       \\
 1.00&-1.54576549& nrl  &-0.97499184&1.752008 &0.83666 &        \\
 1.00&-1.54576963&-4.1401&-0.97499598&1.752008 &0.83666 &0.72811\\
-1.00&-0.54576549&-38.663& 0.02500815&         &0.83666 &0.99836 \\
 2.00&-1.86877182&-3.6984&-1.2129584 &1.524824 &0.69635 &0.59015 \\
-2.00& 0.13111124&-120.62& 0.7869246 &         &0.69635 &0.86803 \\
 3.00&-2.12263028&-3.4913&-1.3955526 &1.375369 &0.60793 &0.51121  \\
 4.00&-2.33530466&-3.3227&-1.5472894 &1.269011 &0.54672 &0.45855 \\
-4.00& 1.66425498&-443.67& 2.4522702 &         &0.54674 &0.69928\\
 8.00&-2.97278069&-2.5994&-1.9998079 &1.027778 &0.41299 &0.34670 \\
10.00&-3.21982390&-2.1806&-2.17498954&0.957089 &0.37519 &0.31551 \\
-10.0& 6.77747928&-2698.9& 7.82231364&         &0.37522 &0.48819\\
\end{tabular}
\tablenotetext[1]{ sign  $B/B_0$ corresponds to the field orientation}
\tablenotetext[2]{{See Ref.\cite{C5}} }
\tablenotetext[3]{{See Ref.\cite{C6}} }
\tablenotetext[4]{In the parentheses brackets the magnitudes Rxy and Qxy
at B/$B_0$ = 0.0 are shown}
\end{table}

\begin{table}
\caption{ $H_2$  total energies(Et),
electronic energies (Es), relativistic correction (Rc),
equilibrium internuclear distance(Re) for $1(1/2_g)^2$
cofiguration of H$_2$, compression magnitude (K) of average
radius in x-y-direction for the total (Rxy) density and small (Qxy)
relativistic component.
In the calculations we have used 432  of basis function
per one relativistic component.
All values are given in a.u.}
\label{table3}
\begin{tabular}{cccccccc}
Config.
\tablenote{
In the nonrelativistic case, the configuration symbols $1(1/2_g)^2$ and
$1(-1/2_g)1(-1/2_u)\,\,$
correspond 1${\sigma^2}_g$ and 1$\sigma_g$1$\sigma_u$
molecular configurations.
}
 &   Et&Rc             &Re    &Shell&  Es&  K(Rxy)& K(Qxy) \\
 &     &$\times10^{-6}$&      &     &    &        &         \\
\hline
\multicolumn{7}{c}{$B/B_0$ = 0.0 }\\
$1(1/2_g)^2$
&-1.1336822&-14.57&1.3866
\tablenote{ See Hartree-Fock calculation by Kolos W. and Roothaan C.C.J.
\cite{C18}
}
&$1(\pm1/2_g)$&-0.5967979&1.0(1.0336)&1.0(1.0906) \\
$1(-1/2_g)1(-1/2_u)\,\,$
&-0.7707848&-17.17&      &1(-1$/2_g$) &-0.9121185 &1.0(0.9030)&1.0(0.9969)  \\
&           &      &      &1(-1$/2_u$) &-0.2091545 &1.0(1.5096)& 1.0(1.3773)  \\
\cline{1-8}
\multicolumn{7}{c}{$B/B_0$ = 0.1 }\\
$1(1/2_g)^2$
&-1.1298513&-14.89&1.3825&1(-1$/2_g$) &-0.6446553 &0.99461 &0.96101 \\
&           &      &      &1(+1$/2_g$) &-0.5446584 &0.99461 &1.03376 \\
\cline{1-8}
\multicolumn{7}{c}{$B/B_0$ = 0.2 }\\
$1(1/2_g)^2$
&-1.1186323&-15.82&1.3719&1(-1$/2_g$) &-0.6882655 &0.98020 &0.92150 \\
&         &        &      &1(+1$/2_g$) &-0.4882717 &0.98019 &1.05855 \\
\cline{1-8}
\multicolumn{7}{c}{$B/B_0$ = 0.35 }\\
$1(1/2_g)^2$
&-1.0894853&-18.35&1.3459&1(-1$/2_g$) &-0.74725115&0.98020 &0.94478 \\
&           &      &      &1(+1$/2_g$) &-0.39726222&0.98019 &1.17371 \\
$1(-1/2_g)1(-1/2_u)\,\,$
&-1.0493226&-13.89&      &1(-1$/2_g$) &-1.06381193&0.96928 &0.94771 \\
&           &      &      &1(-1$/2_u$) &-0.33520119&0.86675 &0.78693 \\
\cline{1-8}
\multicolumn{7}{c}{$B/B_0$ = 0.4 }\\
$1(1/2_g)^2$
&-1.0768969&-19.47&1.3373&1(-1$/2_g$) &-0.7651523 &0.93671 &0.84965 \\
&           &      &      &1(+1$/2_g$) &-0.3651650 &0.93670 &1.07798 \\
$1(-1/2_g)1(-1/2_u)\,\,$
&-1.0806772&-13.76&      &1(-1$/2_g$) &-1.0827429 &0.96148 &0.88676 \\
&           &      &      &1(-1$/2_u$) &-0.3476192 &0.84301 &0.76626 \\
\cline{1-8}
\multicolumn{7}{c}{$B/B_0$ = 0.8 }\\
$1(1/2_g)^2$
&-0.9370938&-33.45&1.2554&1(-1$/2_g$) &-0.8901447 &0.84388 &0.74181 \\
&           &        &    &1(+1$/2_g$) &-0.0901747 &0.84387 &1.04590 \\
\cline{1-8}
\multicolumn{7}{c}{$B/B_0$ = 1.0 }\\
$1(1/2_g)^2$
&-0.8476702&-43.69&1.21696&1(-1$/2_g$)&-0.9430515 &0.80358 &0.70150 \\
&           &      &       &1(+1$/2_g$)& 0.0569073 &0.80357 &1.01730 \\
\cline{1-8}
\multicolumn{7}{c}{$B/B_0$ = 2.0 }\\
$1(1/2_g)^2$
&-0.2894183&-126.9&1.06766&1(-1$/2_g$)&-1.1539132 &0.66011 &0.57082 \\
&           &      &       &1(+1$/2_g$)& 0.8459591 &0.66010 &0.87837 \\
\cline{1-8}
\multicolumn{7}{c}{$B/B_0$ = 4.0 }\\
$1(1/2_g)^2$
& 1.0954315&-454.2&0.898   &1(-1$/2_g$)&-1.4463909&0.51404 &0.44690 \\
&           &     &         &1(+1$/2_g$)& 2.5531496&0.51404 &0.70365 \\
\cline{1-8}
\multicolumn{7}{c}{$B/B_0$ =10.0 }\\
$1(1/2_g)^2$
& 5.9491461&-2726.6&0.68682&1(-1$/2_g$)&-1.9869365&0.35056 &0.31030 \\
&           &      &        &1(+1$/2_g$)&8.0103148&0.35059 &0.48887 \\
\end{tabular}

\end{table}

\begin{table}
\caption{ The comparison of Hydrogen atom compression
magnitudes $K$  for $R_{xy}$  and $R_z$  with relative contraction
of internuclear distance  $D/D_0$  under effect of magnetic field}
\label{table3.1}
\begin{tabular}{dcccc}
 $B/B_0$& $K(R_z)$& $K(R_{xy})$&  $D/D_0$        & $D/D_0$     \\
        &   $H$  &   $H$   &  $H_2^+$        & $H_2$       \\
\tableline
 0.00&  1.0     & 1.0          & 1.0           &  1.0      \\
 0.10&  0.9955  & 0.9914       & 0.9975        &  0.9970   \\
 0.20&  0.9839  & 0.9694       & 0.9904        &  0.9894   \\
 1.00&  0.8632  & 0.7552       & 0.8772        &  0.9137   \\
 2.00&  0.7716  & 0.6070       & 0.7634        &  0.7699   \\
 3.00&  0.7156  & 0.5222       & 0.6886        &           \\
 4.00&  0.6752  & 0.4656       & 0.6354        &  0.6476   \\
 8.00&  0.5845  & 0.3463       & 0.5146        &           \\
10.00&  0.5574  & 0.3134       & 0 4791        &  0.4953   \\
\end{tabular}
\end{table}

\end{document}